\documentclass[12pt]{article}
\usepackage{latexsym}
\usepackage{amsfonts}
\usepackage{graphicx} 
\usepackage{epsfig}

\def\gsim{\mathrel{\rlap {\raise.5ex\hbox{$ > $}}
{\lower.5ex\hbox{$\sim$}}}}
\def\lsim{\mathrel{\rlap {\raise.5ex\hbox{$ < $}}
{\lower.5ex\hbox{$\sim$}}}}

\newcommand{\be}{\begin{equation}}
\newcommand{\ee}{\end{equation}}
\newcommand{\bea}{\begin{eqnarray}}

\newcommand{\eea}{\end{eqnarray}}

\def\gappeq{\mathrel{\rlap {\raise.5ex\hbox{$>$}}
{\lower.5ex\hbox{$\sim$}}}}
\def\lappeq{\mathrel{\rlap{\raise.5ex\hbox{$<$}}
{\lower.5ex\hbox{$\sim$}}}}

\begin{document}

\begin{titlepage}

\begin{flushright}
LAPTH---06\\
CERN-HP-TH/2006-118\\
PNPI----06\\
\end{flushright}

\vspace{0.1in}

\begin{centering}

\Large {\bf{Ternary numbers and algebras}} \\
\vspace{0.1in}

\Large{Alexey Dubrovski}
\footnote {On leave from JINR, Russia}
\Large{and Guennadi Volkov}
\footnote {On leave from PNPI, Russia}

\date{11.1.2006 }

\begin{abstract}
The Calabi-Yau spaces with $SU(n)$ holonomy
can be studied by the algebraic way through  the integer lattice
where  one can construct  the Newton reflexive polyhedra or 
the Berger graphs. Our conjecture is that the Berger graphs can be directly
related with the $n$-ary algebras. To find such algebras we 
study the n-ary generalization of the well-known binary norm 
division algebras, ${\mathbb R}$, ${\mathbb C}$, ${\mathbb H}$,
${\mathbb O}$, which helped to discover the most important "minimal"  binary simple Lie groups,
$U(1)$, $SU(2)$ and  $G(2)$. As the most important example,
 we consider
the case $n=3$, which gives  the ternary generalization of quaternions (octonions), $3^n$, $n=2,3$,
respectively.
The ternary generalization of quaternions  is directly related to the new ternary algebra (group)
which are related to the natural extensions of the  binary $su(3)$ algebra ($SU(3)$ group).  
Using this ternary algebra we found the solution for   the Berger graph:
a  tetrahedron.
\end{abstract}
\end{centering}

\end{titlepage}

\vspace{0.5cm}
\section{Introduction}
Our interest in ternary algebras and symmetries started
from the study of the geometry  based  on the  holonomy principle,
discovered by Berger \cite{Berger}. The $CY_n$ spaces 
with $SU(n)$ holonomy \cite{Calabi, Yau}
have a special interest for us. Our  conjecture \cite{Volkov2, V, LSVV}
is that $CY_3$ ($CY_n$) spaces are related to the ternary ($n$-ary) 
symmetries, 
which are natural generalization
of the binary  Cartan--Killing--Lie symmetries. 

There are some motivations in modern quantum physics and cosmology
why we are searching for new symmetries
beyond Lie algebras/groups.
One of the main reasons  is related with the conjecture that the Standard
Model of quarks and leptons cannot be  solved in terms of  
binary Lie groups. This problem can be formulated as incompleteness of
the Standard Model in terms of binary Lie groups 
\cite{Volkov2,V, LSVV,V3}.
The theory of  superstrings is also based on binary Lie groups, 
in particular  on the
D-dimensional Lorentz group, 
and therefore the description of the Standard Model in the superstrings
approach did not bring us to  success. In our opinion, the main problem 
with the superstring approaches ( also  GUTs, SUGRA) is 
the inadequate external symmetry at
the  string scale, $M_{\rm str} >> M_{\rm SM}$:  
we believe that the D-Lorentz symmetry must 
be generalized. This  is valid also for  GUTs
or SUGRA approaches. 
So far the construction of the quantum
theory of superstrings has not been finished 
and it might be helpul  to know the point limit of superstring theories in 
any dimension $4\leq D \leq 10$. This limit  is known only for 
$D=4$, where
there exist the renormalizable quantum field theories based on the
$D=1+3$ Lorentz group symmetry.  
The  extra uncompactified dimensions make quantum field
theories with Lorentz  symmetry much less credible, since the power 
counting is worse.
A possible way out is to suppose that the propagator is more
convergent than $1/p^2$, such a behaviour can be obtained if we
consider, instead of binary  symmetry algebra, algebras with
higher order relations (\emph{That is, instead of binary
operations such as addition or product of 2 elements, we start with
composition laws that involve at least $n$ elements of the considered
algebra, $n$-ary algebras}). For instance, a  ternary symmetry could  be
related with membrane dynamics.
To solve the Standard Model problems we suggested to generalize their 
external and internal binary symmetries
by addition of ternary symmetries based on the ternary algebras 
\cite{V, LSVV,V3}.
For example,  ternary symmetries seem to give very good possibilities to
overcome  the above-mentioned problems, {\it i.e.} to make the next
progress in understanding of the space-time geometry of our Universe.
We suppose that the new symmetries beyond the well-known binary Lie
algebras/superalgebras could allow us  to build the renormalizable
theories for space-time geometry with dimension $D>4$. It seems
very plausible that using such ternary symmetries will offer a
real possibility to overcome the problems of  quantization of
membranes and  could represent a further progress  beyond string
theories. 

\section{Geometry and algebra of reflexive numbers}

All modern  theories based on the binary Lie algebras
 have a common property since the
algebras/symmetries are related with some invariant quadratic forms.
In all these approaches,  a wide class of  simple
classical Lie algebras is used; the ones
 whose Cartan--Killing classification  contains 
four infinite series 
$A_r=sl(r+1)$, $B_r=so(2n+1)$, $C_r=sp(2r)$, $D_r=so(2r)$, 
and five exceptional algebras ${\it G}_2$, ${\it F}_4$, ${\it E}_6$, 
${\it E}_7$, ${\it E}_8$.  There are several  ways of
studying such classification \cite{Kostr1, Kostr2, FSS, Adams, Baez, Toppan,
Ramond, Traubenberg}. 
We can recall two of them,
one is through the Cartan matrices and Dynkin
graphs, the second  through the theory of numbers and Clifford
algebras. Our interest in  the new $n$-ary algebras and their classification
started from a study of infinite series of $CY_n$ spaces characterized  
by holonomy groups \cite{Berger}. 
More exactly, the $CY_n$ space can be defined as the quadruple
$(M,J,g,\Omega)$, where $(M,J)$ is a complex compact n-dimensional
manifold
of  complex structure $J$, $g$ is a K$\ddot a$hler metrics with $SU(n)$ 
holonomy group, and  
$\Omega_n=(n,0)$ and $\bar \Omega_n=(0,n)$ are non-zero parallel
tensors  which are  the holomorphic volume forms.

A $CY_{n-2}$ space can be realized as an algebraic manifold ${\mathcal M}$
in a weighted projective space
${\rm CP}^{n-1} (\overrightarrow{k})$ 
where the weight vector reads
$\overrightarrow{k} = (k_1, \dots, k_{n})$. This manifold is defined by
\begin{eqnarray}
{\mathcal M} \equiv (\left\{x_1, \dots, x_{n} \right\} \in
{\rm CP}^{n-1} (\overrightarrow{k}):
{\mathcal P} (x_1, \dots, x_{n}) \equiv
\sum_{\scriptsize{\overrightarrow{m}}} c_{\scriptsize{\overrightarrow{m}}}
x^{\scriptsize{\overrightarrow{m}}} = 0 ),
\label{Batyr}
\end{eqnarray}
i.e. {as} the zero locus of a quasi-homogeneous polynomial of
degree $d_k = \sum_{i=1}^{n} k_i$, with the monomials 
$x^{\scriptsize{\overrightarrow{m}}} \equiv x_1^{m_1} \cdots x_{n}^{m_{n}}$. 
The points in
CP$^{n-1}$ satisfy the property of projective invariance
$\left\{x_1, \dots, x_{n}\right\} \approx
\left\{\lambda^{k_1} x_1, \dots, \lambda^{k_{n}} x_{n}\right\}$ leading to the
constraint
$\overrightarrow{m} \cdot \overrightarrow{k} = d_k$.

The classification of $CY_n$ can be done through  
the reflexivity of the weight 
vectors $\overrightarrow{k}$  (reflexive numbers), which can be defined
in terms of the Newton reflexive polyhedra \cite{Bat} or Berger
graphs \cite{V}. The Newton reflexive polyhedra are determined 
by the exponents of the monomials participating in the $CY_n$ equation
\cite{Bat}. The term "reflexive" is related with the mirror duality of 
Calabi--Yau spaces
and the corresponding Newton polyhedra \cite{Bat}.
The Berger graphs can be constructed directly through the reflexive weight 
numbers  $\vec {k}=(k_1,...,k_{n+2})[d_k]$ by thre 
procedure shown in  \cite{V,LSVV}. 
 According to the universal algebraic 
approach \cite{AENV} ( see also \cite{Kurosh, Burris})  
one can find  a section in the  reflexive
polyhedron and, according to the $n$-arity of this algebraic approach,
the reflexive polyhedron can be constructed from 2-, 3-,... Berger graphs. 
It was conjectured that the Berger graphs might correspond 
to $n$-ary Lie algebras \cite{V, LSVV}.
In these articles we tried to decode those Berger graphs by using 
the 
method of the "simple roots" (for illustration, 
see table \ref{BERDET}). 
In this table  some general properties of the 
Berger graphs are presented , which correspond to the   subclass of the 
``simply -laced'' reflexive  numbers (Egyptian numbers).
A simply--laced number $\overrightarrow{k}= (k_1, \cdots, k_n)$
with degree $d_k = \sum_{i=1}^{n} k_i$, is defined such that
\begin{eqnarray}
\frac{d}{k_i} &\in& {\mathbb Z}^+ \, \, {\rm and} \, \, d > k_i.
\end{eqnarray}
For these numbers there is a simple way of constructing the corresponding
affine Berger graphs together with their Coxeter labels \cite{V,LSVV}. The
corresponding Cartan and Berger matrices of these graphs are symmetric. In the 
Cartan case they correspond to the $ADE$ series of simply--laced algebras.
In dimensions $n = 1, 2, 3, $ the Egyptian numbers are
$(1),(1,1),(1,1,1),(1,1,2),(1,2,3)$. For $n=4$, from all 95
reflexive numbers, 14 are simply--laced Egyptian numbers for which 
some general  properties can be illustrated
 (see Table \ref{BERDET}) \cite{LSVV}. 
{\small \begin{table}
\vspace{-3cm}
\hspace{2cm}\begin{tabular}{|l|c|c|c|c|} \hline
${\vec k}_{3,4}^{\rm ext}        $&$ {\rm Rank} $&$h            $&${\rm Casimir} (B_{ii})
$&$ {\rm Determinant}                                                   $  \\
\hline \hline
$(0,1,1,1)[3]      $&$ 6 (E_6)  $&$12         $&$6          $&$3  $  \\
$(0,1,1,2)[4]      $&$ 7 (E_7)  $&$18         $&$8          $&$2  $  \\
$(0,1,2,3)[6]      $&$ 8 (E_8)  $&$30         $&$12         $&$1  $  \\
$(0,0,1,1,1)[3]    $&$ 2_3+10+l $&$18+3(l+1)  $&$9          $&$3^4$  \\
$(0,0,1,1,2)[4]    $&$ 2_3+13+l $&$32+4(l+1)  $&$12         $&$4^3$  \\
$(0,0,1,2,3)[6]    $&$ 2_3+15l  $&$60+6(l-1)  $&$18         $&$6^2$  \\ \hline
$(0,1,1,1,1)[4]    $&$ 1_3+11   $&$28         $&$12         $&$16 $  \\
$(0,2,3,3,4)[12]   $&$ 1_3+12   $&$90         $&$36         $&$ 8 $  \\
$(0,1,1,2,2)[6]    $&$ 1_3+13   $&$48         $&$18         $&$ 9 $  \\
$(0,1,1,1,3)[6]    $&$ 1_3+15   $&$54         $&$18         $&$ 12$  \\
$(0,1,1,2,4)[8]    $&$ 1_3+17   $&$80         $&$24         $&$ 8 $  \\
$(0,1,2,2,5)[10]   $&$ 1_3+17   $&$100        $&$30         $&$ 5 $  \\
$(0,1,3,4,4)[12]   $&$ 1_3+17   $&$120        $&$36         $&$ 3 $  \\
$(0,1,2,3,6)[12]   $&$ 1_3+19   $&$132        $&$36         $&$ 6 $  \\
$(0,1,4,5,10)[20]  $&$ 1_3+26   $&$290        $&$60         $&$ 2 $  \\ \hline
$(0,1,1,4,6)[12]   $&$ 1_3+24   $&$162        $&$36         $&$ 6 $  \\
$(0,1,2,6,9)[18]   $&$ 1_3+27   $&$270        $&$54         $&$ 3 $  \\
$(0,1,3,8,12)[24]  $&$ 1_3+32   $&$420        $&$72         $&$ 2 $  \\
$(0,2,3,10,15)[30] $&$ 1_3+25   $&$420        $&$90         $&$ 4 $  \\
$(0,1,6,14,21)[42] $&$ 1_3+49   $&$1092       $&$126        $&$ 1 $  \\
\hline
\end{tabular}
\caption{Rank, Coxeter number $h$, Casimir depending on $B_{ii}$ and 
determinants
for the non--affine exceptional Berger graphs. The maximal Coxeter labels
coincide with the degree of the corresponding reflexive simply--laced vector.
The determinants in the last
column for the infinite series (0,0,1,1,1)[3], (0,0,1,1,2)[4] and 
(0,0,1,2,3)[6]
are independent from the number $l$ of internal binary $B_{ii}=2$ nodes.
The numbers $1_3$ and $2_3$ denote the number of nodes with $B_{ii}=3$.}
\label{BERDET}
\end{table}}

To get further  progress in  the  solution of the Berger graphs by method 
of "simple roots" we
should construct  the $n$-ary analogue of the $su(2)/u(1)$ node,   
which was the key-stone in  the 
Killing--Cartan--Lie classification \cite{FSS, Adams}.
Therefore one should search for such  "minimal" $n$-ary algebras. 
To do this   we would like to 
use the  ideas  coming  from the theory of the binary  norm division algebras,
${\mathbb R}$, ${\mathbb C}$, ${\mathbb H}$, ${\mathbb O}$.
 But to define the complete decision we must 
find the real examples of the "minimal" 
simple algebras like  $su(2)$ and  $g(2)$,
which are directly related with quaternions and octonions, respectively.
Therefore we concentrate on  searching for the $n$-arity division algebras.

\section{Division algebras and Lie algebras} 
Now we should briefly recall  the four norm division algebras
${\mathbb R}$,${\mathbb C}$,${\mathbb H}$,${\mathbb O}$ 
\cite{Hamilton, Cayley, Kostr1, Kostr2, Baez,Toppan}. 
An algebra $A$ will be a vector space that is
equipped with a bilinear map $f:A \times A \rightarrow A$ called by
multiplication and a nonzero element $1 \in A$ called the unit, such
that $f(1,a)=f(a,1)=a$. These algebras admit an anti-involution (or  
 conjugation) 
$(a^*)^*=a$ and $ (ab)^*={b}^*  {a}^*$.
A norm division algebra is an algebra $A$ that
is also a normed vector space with $N(ab)=N(a)N(b)$.
Such algebras exist only for $n=1,2,4,8$ dimensions where 
the following identities can be obtained:
\begin{eqnarray}
(x_1^2+...+x_n^2)(y_1^2+...+y^2)=(z_1^2+...+z_n^2)
\end{eqnarray}
The doubling process, which is known as the Cayley-Dickson process,
forms  the sequence of divison algebras
\begin{eqnarray}
{\mathbb R} \rightarrow {\mathbb C} \rightarrow {\mathbb H}
 \rightarrow { \mathbb O}.
\end{eqnarray}
Note  that next algebra is not a division algebra.
So $n=1$ ${\mathbb R}$ and $n=2$ ${\mathbb C}$ these algebras 
are  the  commutative associative normed division algebras.
The quaternions,  ${\mathbb H}$, $n=4$ form the  non-commutative
and associative norm division algebra. 
The octonion algebra  $n=8, {\mathbb O}$ is  an 
non-associative alternative algebra.
If the discovery of complex numbers took a long period about some
centuries years, the discovery of quaternions and octonions was made
in  a short time, in the middle of the ${\rm XIX}$ century by 
W. Hamilton \cite{Hamilton}, and by
J. Graves and  A.Cayley \cite{Cayley}. 
The complex numbers, quaternions
and octonions can be presented in the general form:

\begin{eqnarray}
\hat q= x_0 e_0+x_pe_p,\qquad  \{x_0,\,x_p\} \in {\mathbb R},
\end{eqnarray} 
where $p=1$ and $e_1 \equiv {\bf i}$   for complex numbers ${\mathbb C}$,
$p=1,2,3$ for quaternions ${\mathbb H}$, and $p=1,2,...,7$ for
${\mathbb O}$. The $e_0$ is as unit and all $e_p$ are imaginary units
with conjugation $\bar {e}_p=-e_0$.
 For quaternions we have the main  relation 
\begin{eqnarray}
e_me_p=-\delta_{mp} +f_{mpl}e_l,
\end{eqnarray}
where $\delta_{mp}$ and $f_{mpl}\equiv \epsilon_{mpl}$ are 
the well-known Kronecker and Levi-Cevita tensors, respectively.
For octonions the completely antisymmetric tensor
$f_{mpl}=1$ for the following seven triple associate cycles:
\begin{equation}
\{mpl\}=\{123\},\{145\},\{176\},\{246\},\{257\},\{347\},\{365\}.
\end{equation}

There are also 28 non-associate cycles.
Each triple accociate cycle corresponds to a quaternionic subalgebra.
These algebras have a very close link with geometry.
For example, the unit elements  $x^2=1, x \in {\mathbb R}$,
$|\hat q|=x_0^2+x_1^2=1$ in $ {\mathbb C}_1$,  
$|\hat q|=x_0^2+x_1^2+x_2^2+x_3^2=1$ in ${\mathbb H}_1$,  
$|\hat q|=x_0^2+x_1^2+...+x_7^2=1$ in ${\mathbb O}_1$,
define the spheres, $S^0$,  $S^1$, $S^3$,
$S^7$, respectively. The complex and quaternionic unit elements
have the $U(1)$ and $SU(2)$ group properties, 
respectively. The $A$ series contains the complex rotations in the
unit circle, $S^1$, and $S^1$ is a Lie group.
The $B$ and $C$ groups both contain the quaternion rotations on the
unit sphere $S^3$, and $S^3$ is a Lie group.
The $D$ series contain the Lorentz group in $D=3+1$, which consists of two 
copies of $S^3$-3-rotations and 3-boosts.
The exceptional groups do not include $S^7$ as a Lie group.
Thus $S^7$ is the only unit sphere in a division algebra that is not a 
Lie group. The reason  is that   
the octonions  are not associative; their  
associator is 
\begin{eqnarray}
\{e_m,e_p,e_l\}=(e_me_p)e_l-e_m(e_pe_l)=f_{mplk}e_k,
\end{eqnarray}
where the tensor $f_{mplk}$ is  completely antisymmetric and it 
is non-zero for the following  seven 4-cycles:

\begin{equation}
\{mplk\}=\{4567\},\{2367\},\{2345\},\{1357\},\{1346\},\{1256\},\{1247\}.
\end{equation}
It is also the case  when three elements $\{e_a,e_b,e_c\}$ are not in the same three
associate cycles, for example, $(e_1e_5)e_7-e_1(e_5e_7)=2e_3$.
   Note that the octonions form the alternative algebra since
the alternative condition, $\{a,b,c\}+\{c,b,a\}=0$, is always valid.
The octonions are directly linked to the five exceptional groups $G(2), F(4), E(6),E(7),E(8)$
\cite{Adams, Baez}. The automorphism of octonions is $G(2)$.

\section{Nambu--Filippov ternary  algebras}

The new n-ary    symmetries can be related to algebras that are
based on  the generalization of the Lie binary commutation
relation
\begin{equation}
 [x,y]=xy-yx
\end{equation}
 by the ternary  commutations relations
\begin{equation}
[x_1,x_2,...,x_n]=(-1)^{\tau(\sigma)}
[x_{\sigma(1)},x_{\sigma(2)},...,x_{\sigma(n)}],
\end{equation}
 where $\sigma$ runs
over the symmetric group $S_n$  and the number
$\tau(\sigma)$ equals  0 or 1, depending on the parity of the
permutation $\sigma$ 
(see \cite{Kurosh,Nambu,Filippov,CIZ,Takhtajan,Gned,Hanlon,Michor,Ibanez,Kerner,Abramov,Nakanishi,
Bremner,Bremner2, Burris}).

 More exactly, a ternary Lie algebra is defined by a ternary  
antisymmetric operation
$A\times A\times A \rightarrow A$ with the 
Jacobi-like identity \cite{Filippov}. Fillipov considered 
the $n$-ary generalizations of  Lie algebras ($n>2$). 
The most simple example of this  Lie algebra can be 
the $n$-vector product of the 
$(n+1)$- dimensional vectors $\vec x_1,...,\vec x_n$ which is equal 
to the following determinant:
\begin{eqnarray}
[\vec x_1,...,\vec x_n]_n&=&\vec x_1 \times\vec x_2 \times \ldots \times
\vec x_n= \nonumber\\
&& \left\arrowvert{
\begin{array}{ccccc}
x_{11}        &x_{12}       &\ldots      &x_{1n}      &\vec e_1     \\
x_{21}        &x_{22}       &\ldots      &x_{2n}      &\vec e_2      \\
\ldots        &\ldots       &\ldots      &\ldots      &\ldots      \\
x_{(n+1) 1}   &x_{(n+1)2}   &\ldots      &x_{(n+1)n}  &\vec e_{n+1} \\
\end{array}}
\right\arrowvert ,
\end{eqnarray} 
where $(x_{1 l},\ldots x_{(n+1)l} )$ are the coordinates of the 
vector $\vec x_l$ and $\{e_l\}$ is the orthonormal basis.

The other simple example is the algebra  of polynomials of $n$-variables
$x_1,\ldots x_n$ where    n-ary operation is the 
functional Jacobian:
\begin{equation}  
[f_1,...,f_n]_n= 
{\rm det} ||\frac{\partial (f_1,\ldots, f_n)}{\partial (x_1,\ldots x_n)}||.
\end{equation}
The $n$-ary Poisson-like structure of the determinat
have been used for the   
generalization of the classical Hamiltonian mechanics
in which the binary Poisson 
bracket can be  replaced by  ternary \cite{Nambu} or  
by  n-ary brackets \cite{Filippov, Takhtajan}, respectively.

There  exist  several examples of multi-Hamiltonian systems possessing 
dynamical or hidden symmetries, which can be realized 
within the generalized Nambu--Hamiltonian mechanics using the $n$-ary 
Nambu-Poisson brackets ($n>2$).
Among such systems one can consider the $SO(4)$ Kepler problem
\cite{Chatter}. The well-known Kepler Hamiltonian is $H=\vec p^2/2-1/r$,
where $r=\sqrt{x_1^2+x_2^2+x_3^2}$. Such a Hamiltonian   
has the $SO(3)$ rotational symmetry giving the orbital angular
momentum $\vec L=\vec r \times \vec p$ as an integral of motion.
This integral of motion with a Hamiltonian implies that the 
orbit lies in 2-dimensional plane, but cannot explain why it is
closed. Laplace discovered 
the other hidden $SO(3)$ symmetry, which gives the additional  integral
of motion $\vec A= \vec p \times \vec L - \vec r/r$.
The Kepler problem was naturally solved in Nambu Hamiltonian mechanics,
in which the equations of motion are given by n-Poisson--Nambu bracket:
\begin{equation}
\frac{df}{dt}=[H_1,..,H_5,f].
\end{equation}

Other n-ary analogues of Lie algebras have also been considered by 
Bremner \cite{Bremner, Bremner2}.   
Bremner found the minimal (Jacobi) identity in the totally associative 
case, which has degree 7. One calls algebra $A$ totally associative if
it satisfies the polynomial identities

\begin{eqnarray}
t_i(a_1,a_2,...,a_{2n-1})=a_1...(a_i...a_{i+n-1})...a_{2n-1}-
a_1...(a_{i+1}...a_{i+n})...a_{2n-1}=0,
\end{eqnarray}
for $1 \leq i \leq n-1$.
The minimal identity in the partially associative case, which has
degree 5, was found by Gnedbaye \cite{Gned}; $A$ is called partially associative if it 
satisfies the polynomial identity

\begin{eqnarray}
p(a_1,...,a_{2n-1})=\sum_{i=1}^{i=n} (-1)^{(n-1)i}
a_1...a_{i-1}(a_i...a_{i+n-1})a_{i+n}...a_{2n-1}=0.
\end{eqnarray}

Also the ternary symmetries  have been intensively  discussed 
in quantum physics \cite{Kerner, Abramov}, in conformal field theories 
\cite{CIZ}.

One of the best ways of studing the n-ary algebras/symmetries is through 
the n-ary generalizations of Clifford algebras.
The binary Clifford algebra is an associative
algebra, which contains and is generated by a vector space having a
quadratic form: 
$e_ke_l+e_le_k=2 g_{kl}$;
$\{e_a\}$ is the  basis of the  vector space and the signature of this
space is determined by the  metric
$g_{kl}=diag(-1,...,-1_t,+1_,...,1_s)$.

The binary Clifford algebras are closely
related with the study of the rotation groups of multidimensional 
spaces. The $1_t+3_s$ Minkowski space can be described by the
quadratic form $x^2+y^2+z^2-c^2 t^2$, which remains invariant under
a general Lorentz group. The general Lorentz group $O(3,1)$ consists of  a
proper orthochonous Lorentz group $O_0(3,1)$ and three reflections
(discrete transformations) $P, T, PT$, where $P$ and $T$ are space and 
time reversal. In the general case of the real space $R^{s,t}$, the
orthogonal group $O(s,t)$ is represented by the semidirect product 
of a connected component $O(s,t)_0$ and a discrete subgroup $\{1, P,
T, PT \}$. The double covering of the orthogonal group $O(s,t)$ is a
Clifford group $Pin(s,t)$, which can be completely constructed within a 
Clifford algebra, {\it i.e.}  $Pin(s,t) \subset {\rm Cl}_{s,t} $ 
\cite{Varlamov}. The discrete symmetries  are represented by
fundamental automorphisms of the Clifford algebras,
{\it i.e. } $\{1,P,T, PT \} \approx {\rm Aut} ({]rm Cl})$. 
In contrast with the transformations $P,T, PT$, the operation 
${\rm C}$ of charge conjugation is represented by a pseudoautomorphism 
$ A \rightarrow \bar A$.  An extended  automorphism group $Ext (Cl)$
is isomorphic to a $CPT$ group $\{1,P,T,PT,C, CP,CT,CPT\}$.  
The n-ary Clifford algebras are related to the generalizations of
orthogonal groups.

{\section{The ternary generalization of quaternions
and the tripling Cayley-Dickson method}
 
We would now like to discuss the doubling Cayley--Dickson method
and generalize its to construct the ternary form division algebras.
The complex numbers are 2-dimensional algebra with basis
$e_0$ and $e_1 \equiv {\bf i}$,
\begin{equation}
{\mathbb C}={\mathbb R} \oplus {\mathbb R} e_1,
\end{equation} 
where $e_0^2=e_0$, $e_1e_0=e_0e_1=e_1$ and
$e_1$ is the imaginary unit, ${\bf i}^2=-e_0$. 
Considering one additional basis imaginary unit element $e_2\equiv {\bf j}$ 
in the Cayley--Dickson doubling process, we can obtain the quaternions:
\begin{equation}
{\mathbb H}={\mathbb C} \oplus  {\mathbb C} j.
\end{equation}

By the Cayley--Dickson method,
 if $a, b, c, d  \in X$, when $(a, b), (c, d) \in X^2$.
And the product is defined in $X^2$ through
\begin{equation}
 (a, b) (c, d) = (ad - c^*b, bc^* + da)
\end{equation}

This means that quaternions can be considered as a pair of complex numbers:
\begin{equation}
q=(a+ {\bf i}b)+j( c+ {\bf i} d),
\end{equation}
where 
\begin{equation}
j( c+ {\bf i} d)= (c+ \bar {\bf i} d) j=(c - {\bf i}d ) j.
\end{equation}
so, that we can see that $ {\bf i} {\bf j}=-{\bf j}{\bf i}={\bf k}$.

The quaternions 
\begin{equation}
q=x_0e_0 +x_1e_1+x_2e_2+x_3e_3, \qquad q\in {\mathbb H}, 
\end{equation}
produce over ${\mathbb R}$  a 4-dimensional norm division algebra, 
where the fourth imaginary unit
$e_3=e_1e_2 \equiv {\bf k}$ appears.  The main multiplication rules of all these four 
elements  are the following:
\begin{eqnarray}
&&{\bf i}^2={\bf j}^2={\bf k}^2=-1 \nonumber\\
&&{\bf ij }={\bf k} \qquad{\bf ji }=-{\bf k}. \nonumber\\ 
\end{eqnarray}
All other identities can be obtained from cyclic permutations
of ${\bf i,j,k}$.
The imaginary quaternions ${\bf i,j,k}$ produce the $su(2)$ algebra.
The matrix realization of quaternions has been done through the Pauli
matrices:
\begin{eqnarray}
\sigma_0, \,\,{\bf i} \sigma_1, {\bf i} \sigma_2, {\bf i} \sigma_3.
\end{eqnarray}

The unit quaternions $q=a  {\bf 1}+b {\bf i}+c {\bf j}+d {\bf k}
\in H_1$, $q \bar q=1$, produce the $SU(2)$ group:
\begin{eqnarray}
q \bar q=a^2+b^2+c^2+d^2=1, \,\, \{a,b,c,d\}\in S^3, \, S^3 \approx SU(2).
\end{eqnarray}

Simililarly,  continuing  the Cayley--Dickson doubling process,  
we can build the octonions:
\begin{eqnarray}
{\mathbb O} ={\mathbb H} \oplus {\mathbb H} {\bf l}, \nonumber\\
\end{eqnarray}
where we introduced the new basis element ${\bf l} \equiv e_4$. As a result of this process, 
the basis $\{1,{\bf i}, {\bf j}, {\bf k}\}$ of ${\mathbb H}$
is complemented to a basis $\{ {\bf 1}=e_0, {\bf i}=e_1, {\bf j}=e_2, {\bf k}=e_3=e_1e_2, 
{\bf l}=e_4, {\bf il}=e_5=e_1e_4, {\bf jl}=e_6=e_2e_4, {\bf kl}=e_7=e_3e_4   \}$ 
of ${\mathbb O}$:
\begin{eqnarray}
{\it o} = x_0e_0+x_1e_1+x_2e_2+x_3e_3+x_4e_4+x_5e_5+x_6e_6+x_7e_7,
\end{eqnarray}
where we can see the  following seven associative cycle triples:
\begin{eqnarray}
&&\{123:e_1e_2=e_3\}, \,\{145:e_1e_4=e_5\},\, \{176:e_1e_7=e_6\},\, \{246:e_2e_4=e_6\},  \nonumber\\
&&\{257:e_2e_5=e_7\}, \,\{347:e_3e_4=e_7\},\, \{365:e_3e_6=e_5\}.
\end{eqnarray}

The doubling process can consequently produce new algebras:

\begin{eqnarray}
{\mathbb R} \rightarrow {\mathbb C} \rightarrow {\mathbb H} \rightarrow 
{\mathbb O} \rightarrow {\mathbb S} \rightarrow  \ldots,
\end{eqnarray}
but just the first four from this list are the norm division alternative algebras.

In order find new division algebras, we can just relax
some  constraints.
To do this, let us consider two basic elements, $q_0$, $q_1$, with the following constraints:
\begin{eqnarray}
q_1 \cdot q_0=q_0 \cdot q_1=q_1,\,\,\,\, q_1^3=q_0.\nonumber\\ 
\end{eqnarray} 
In this case a new element can be introduced, as  $q_2 =q_1^2=q_1^{(-1)}$,
{\it i.e.} $q_2q_1=q_1q_2=q_0$. 

From these three elements a new
field ${\mathbb TC}$ can be built:
\begin{eqnarray}
{\mathbb TC } = {\mathbb R} \oplus {\mathbb R} q_1  \oplus {\mathbb R} q_1^2,
\end{eqnarray}
with the new numbers  
\begin{eqnarray}
\hat z=x_0 q_0 + x_1 q_1 + x_2 q_2, \qquad x_i \in {\mathbb R}, \qquad i=0,1,2,
\end{eqnarray}
which are the ternary generalization  of the complex numbers. 
Note that the  multicomplex  numbers, 
${\mathbb MC_n}=\{x=\sum_{i=0}^{i=n-1} x_iq^i, q^n=-q_0,x_i \in {\mathbb R}\}$
have been suggested  in  articles
\cite{Fleury, Fleury2, Traubenberg}.

Let us define consequent two operations of the ${\mathbb Z}_3$-conjugation:
\begin{eqnarray}
\bar q_1=j q_1, \qquad {\bar {\bar q}}_1 =j^2 q_1,
\end{eqnarray}
where $j=\exp {(2 {\bf i} \pi)/3}$.
Since $q_2=q_1^2$ we can easily obtain 
\begin{eqnarray}
{\bar q}_2=j^2 q_2, \qquad {\bar {\bar q}}_2 =j q_2.
\end{eqnarray}
One can also define a ${\mathbb Z}_2$-conjugation which transforms $q_1 \rightarrow q_1^2=q_2$ and $q_2 \rightarrow q_2^2=q_1$, 
{\it i.e.}  $q_1^*=q_2$ and $q_2^*=q_1$. 
Two ${\mathbb Z}_3$-conjugation operations can be applied, respectively: 
\begin{eqnarray}
&&       {\bar {\hat z}} = x_0 q_0 + x_1 j   q_1 + x_2 j^2q_2, \nonumber\\
&&{\bar  {\bar {\hat z}}}= x_0 q_0 + x_1 j^2 q_1 + x_2 j  q_2. \nonumber\\
\end{eqnarray}
We now introduce the cubic form:
\begin{eqnarray}
\langle{\hat z}\rangle={\hat z} {\bar {\hat z}} {\bar {\bar {\hat z}}}=x_0^3+x_1^3+x_2^3-3x_0x_1x_3,
\end{eqnarray}

And also easily check the following relation: 
\begin{eqnarray}
\langle{\hat z}_1{\hat z}_2\rangle=\langle{\hat z}_1\rangle\langle{\hat z}_2\rangle,
\end{eqnarray}
which indicates  the group properties of the    ${\mathbb TC}$ numbers.
More exactly, the unit ${\mathbb TC}$ numbers produce the  Abelian ternary group.
According to the ternary analogue of the Euler formula, the following ternary 
complex  functions \cite{Fleury, Fleury2,Traubenberg} can be constructed:

\begin{eqnarray}
\Psi =\exp{(q_1 \phi_1 + q_2\phi_2) },\qquad
\psi_1=\exp{(q_1 \phi_1) }, \qquad 
\psi_2=\exp{(q_2 \phi_2) },  
\end{eqnarray}
where $\phi_i$ are the  group  parameters.
For  the functions $\psi_i$, $i=0,1,2$, {\it i.e.} we have the following analogue of Euler, formula:
\begin{eqnarray}
\Psi&=&\exp{(q_1\phi+q_2 \phi_2)}  =f   q_0+ g   q_1+h   q_2,\nonumber\\
\psi_1&=&\exp{(q_1\phi)}           =f_1 q_0+ g_1 q_1+h_1 q_2,\nonumber\\
\psi_2&=&\exp{(q_2\phi)}           =f_2 q_0+ h_2 q_1+g_2 q_2,\nonumber\\
\end{eqnarray}
Consequently, e can now
introduce the conjugation operations for these functions. For example, for $\psi_1$ we can get:
\begin{eqnarray}
{\bar \psi}_1&=&\exp {({\bar q_1} \phi)}= 
\exp{(j \cdot q_1\phi)}=f q_0+ jg q_1+ j^2h q_2,\nonumber\\
{\bar {\bar \psi}}_1&=&\exp{({\bar {\bar q}}_1 \phi)}=
\exp{(j^2 \cdot q_1\phi)}=f_1 q_0+ j^2 g_1 q_1+j h_1 q_2,\nonumber\\
\end{eqnarray}
with the following constraints:

\begin{equation}
\psi_1 {\bar \psi}_1 {\bar{\bar \psi}}_1
= \exp{( q_1\phi)}\exp{(j \cdot q_1\phi)}\exp{(j^2 \cdot q_1\phi)}=q_0,
\end{equation}
which gives us the following link between   the functions, $f,g,h$:
\begin{equation}
f_1^3+g_1^3+h_1^3-3 f_1g_1h_1=1.
\end{equation}
This surface (see figure \ref{fig:pic1s}, appendix: $x^3+y^3+u^3-3xyu=1$ and \cite{KernerChessboard}) is a ternary analogue of the $S^1$ circle and it is related with the ternary Abelian group, $TU(1)$.

\begin{figure}[htpb]
  \begin{center}
    \mbox{\epsfig{file=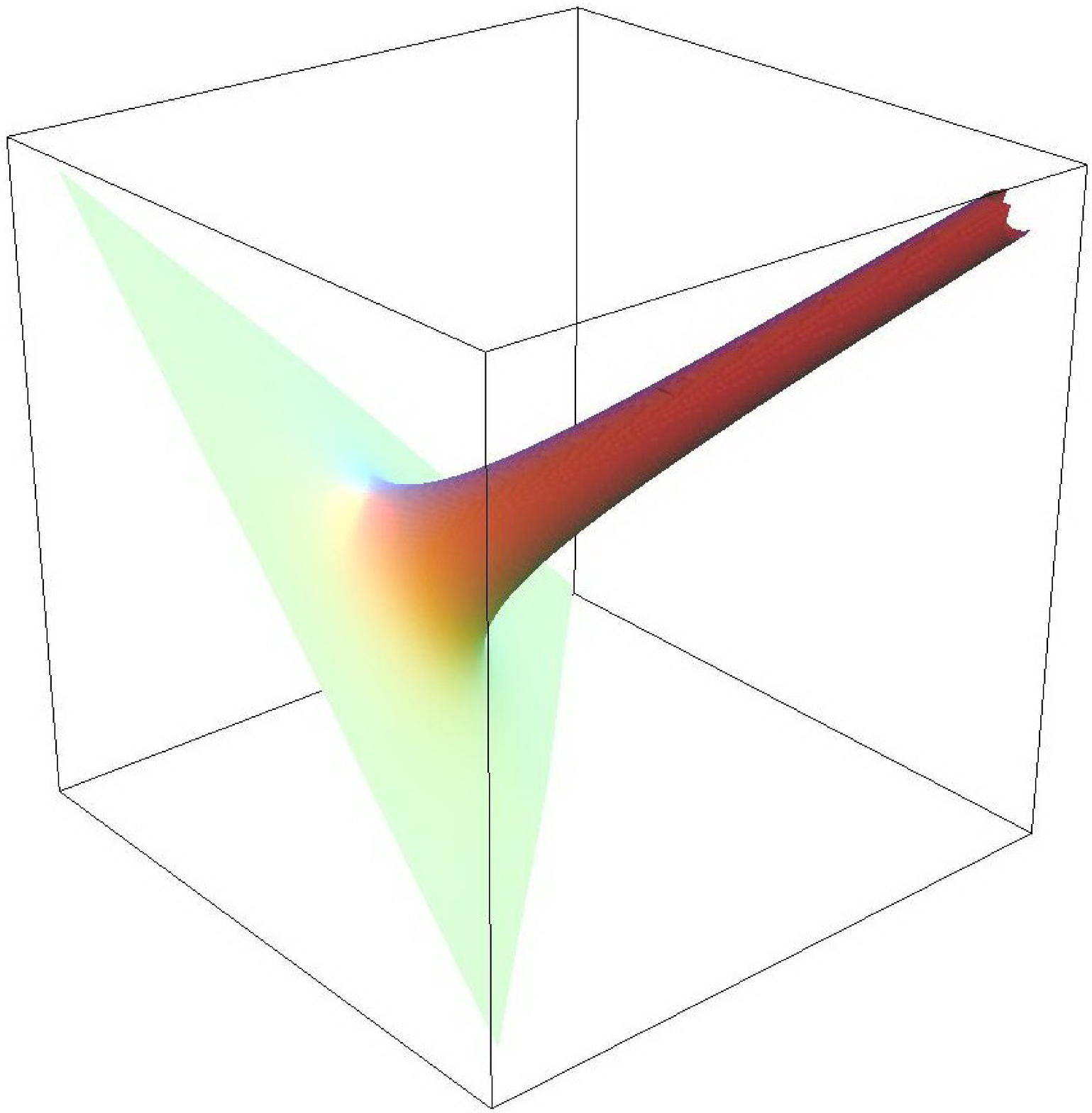,height=13.36cm,width=14cm}}
  \end{center}
  \caption{The surface for $f_1^3+g_1^3+h_1^3-3 f_1g_1h_1=1$}
  \label{fig:pic1s}
\end{figure}

At the next step, we  can  construct  
the analogue of binary quaternions ${\mathbb H}$,
based on the three elements $q_0,q_1,q_7$. 
Our new constraint
$q_1^3=q_7^3=q_0$, we can introduce the following six new products from  $q_1$ and $q_7$:
$q_1^2, q_7^2, q_1q_7$, $q_1^2q_7,q_7^2q_1$ and $q_1^2q_7^2$. 
More exactly,  to get the new $3^2$ ternary  ${\mathbb H}$ numbers,
we suggest the tripling  Cayley-Dickson process starting from ${\mathbb TC}$: 

\begin{eqnarray}
{\mathbb H} = {\mathbb TC} \oplus {\mathbb TC} q_7  \oplus {\mathbb TC} q_7^2,
\end{eqnarray}
or more exactly
\begin{eqnarray}
Q&=&      (x_0q_0 +    x_1q_1 +     x_2q_1^2)
  + q_7   (y_0q_0 +    y_1q_1 +     y_2q_1^2)
  + q_7^2 (z_0q_0 +    z_1q_1 +     z_2q_1^2)                              \nonumber\\
 &=&      (x_0q_0 +    x_1q_1 +     x_2q_4  )
  +       (y_0q_7 + j  y_1q_2 + j   y_2q_6  )
  +       (z_0q_8 +    z_1q_3 + j^2 z_2q_5  )                              \nonumber\\
 &=&      (x_0q_0 +    y_0q_7 +     z_0q_8  )
  +       (x_1q_1 +    y_1q_2 +     z_1q_3  )
  +       (x_2q_4 + j  y_2q_6 + j^2 z_2q_5),                               \nonumber\\
\end{eqnarray}
where we defined the new unit elements 
$q_1^2=q_4,    q_7q_1=q_2, q_7^2=q_8$,
$q_7^2q_1=q_3, q_7 q_1^2=j q_6$, 
$q_7^2q_1^2=j^2 q_5$ with $q_a^3=q_0$.

To find the table of multiplication (see table \ref{MUL}) of 
all $q_a$ basis elements 
we can recall the  identity between three
unit imaginary elements, $e_1,e_2,e_3$, in binary quaternion algebra:
\begin{eqnarray}
&&e_1 e_2 e_3\,=\,e_2 e_3 e_1\,=\,e_3e_1e_2\,=\,-\,e_0, \nonumber\\
&&e_3 e_2 e_1\,=\,e_2 e_1 e_3\,=\,e_1e_3e_2\,=\,+\, e_0.
\end{eqnarray}
Since we suppose that  the ${\mathbb H}$ numbers are the ternary 
generalizations
of quaternions  we can start from the following  
triple identities for $q_1,q_2,q_3$:

\begin{eqnarray}
&&q_1q_2q_3=q_2q_3q_1=q_3q_1q_2=j^2q_0, \nonumber\\ 
&&q_3q_2q_1=q_2q_1q_3=q_1q_3q_2=jq_0, \nonumber\\ 
\end{eqnarray}
where $j=\exp(2\pi {\bf i}/3)$.

Introducing the new elements
\begin{eqnarray}
q_1^2=q_4, \qquad q_2^2=q_5, \qquad q_3^2=q_6,
\end{eqnarray}
we can immediately get the following triple identities:

\begin{eqnarray}
&&q_4q_5q_6=q_5q_6q_4=q_6q_4q_5=j^2q_0, \nonumber\\
&&q_6q_5q_4=q_5q_4q_6=q_4q_6q_5=jq_0.   \nonumber\\
\end{eqnarray}
We suggest the following  $24$ commutation relations:
\begin{eqnarray}
\begin{array}{ccc}
q_1q_2=j  q_2q_1, &q_2q_3=j  q_3q_2, & q_3q_1=j  q_1q_3 \\
q_4q_5=j  q_5q_4, &q_5q_6=j  q_6q_5, & q_6q_4=j  q_4q_6 \\
q_5q_1=j  q_1q_5, &q_6q_2=j  q_2q_6, & q_4q_3=j  q_3q_4 \\
q_6q_1=j^2q_1q_6, &q_5q_3=j^2q_4q_2, & q_4q_2=j^2q_5q_3 \\
q_1q_7=j  q_7q_1, &q_2q_7=j  q_7q_2, & q_3q_7=j  q_7q_3 \\
q_1q_8=j^2q_8q_1, &q_2q_8=j^2q_8q_2, & q_3q_8=j^2q_8q_3 \\
q_7q_4=j  q_4q_7, &q_7q_5=j  q_5q_7, & q_7q_6=j  q_6q_7 \\
q_8q_4=j^2q_4q_8, &q_8q_5=j^2q_5q_8, & q_8q_6=j^2q_6q_8 \\
\end{array}
\end{eqnarray}
and $4$ commuting pairs:
\begin{eqnarray}
q_1q_4=q_4q_1,\, q_2q_5=q_5q_2,\,q_3q_6=q_6q_3,\,q_7q_8=q_8q_7.
\end{eqnarray}

\begin{table}
\centering
\caption{ \it  The binary multiplication relations}
\label{MUL}
\vspace{.05in}
\begin{tabular}{|c||c|c|c||c|c|c||c|c|c||}
\hline
$     N   
$&$      q_1       $&$     q_2      $&$     q_3  
$&$      q_4       $&$     q_5      $&$     q_6 
$&$      q_7       $&$     q_8      $&$     q_0
$\\ \hline\hline  
$        q_1   
$&$      q_4       $&$ j^2 q_6      $&$  j   q_5
$&$      q_0       $&$     q_8      $&$      q_7
$&$  j   q_2       $&$ j^2 q_3      $&$      q_1
$\\ \hline 
$        q_2
$&$  j   q_6       $&$     q_5      $&$  j^2 q_4
$&$      q_7       $&$     q_0      $&$      q_8
$&$  j   q_3       $&$ j^2 q_1      $&$      q_2
$\\ \hline 
$        q_3
$&$  j^2 q_5       $&$ j   q_4      $&$      q_6
$&$      q_8       $&$     q_7      $&$      q_0
$&$  j   q_1       $&$ j^2 q_2      $&$      q_3
$\\ \hline\hline 
$        q_4
$&$      q_0       $&$ j^2 q_7      $&$  j   q_8
$&$      q_1       $&$ j^2 q_3      $&$  j   q_2
$&$      q_6       $&$     q_5      $&$      q_4
$\\ \hline 
$        q_5
$&$  j   q_8       $&$     q_0      $&$  j^2 q_7
$&$  j   q_3       $&$     q_2      $&$  j^2 q_1
$&$      q_4       $&$     q_6      $&$      q_5
$\\ \hline 
$        q_6
$&$  j^2 q_7       $&$ j   q_8      $&$      q_0
$&$  j^2 q_2       $&$ j   q_1      $&$      q_3
$&$      q_5       $&$     q_4      $&$      q_6
$\\\hline\hline
$        q_7
$&$      q_2       $&$     q_3      $&$      q_1
$&$  j   q_6       $&$ j   q_4      $&$   j  q_5
$&$      q_8       $&$     q_0      $&$      q_7
$\\ \hline 
$        q_8
$&$      q_3       $&$     q_1      $&$      q_2
$&$  j^2 q_5       $&$  j^2q_6      $&$  j^2 q_4
$&$      q_0       $&$     q_7      $&$      q_8
$\\ \hline 
$        q_0  
$&$      q_1       $&$     q_2      $&$      q_3  
$&$      q_4       $&$     q_5      $&$      q_6 
$&$      q_7       $&$     q_8      $&$      q_0
$\\ \hline\hline 
\end{tabular}
\end{table}

\begin{figure}
        \centering
                \includegraphics[type=eps,ext=.eps,read=.eps]{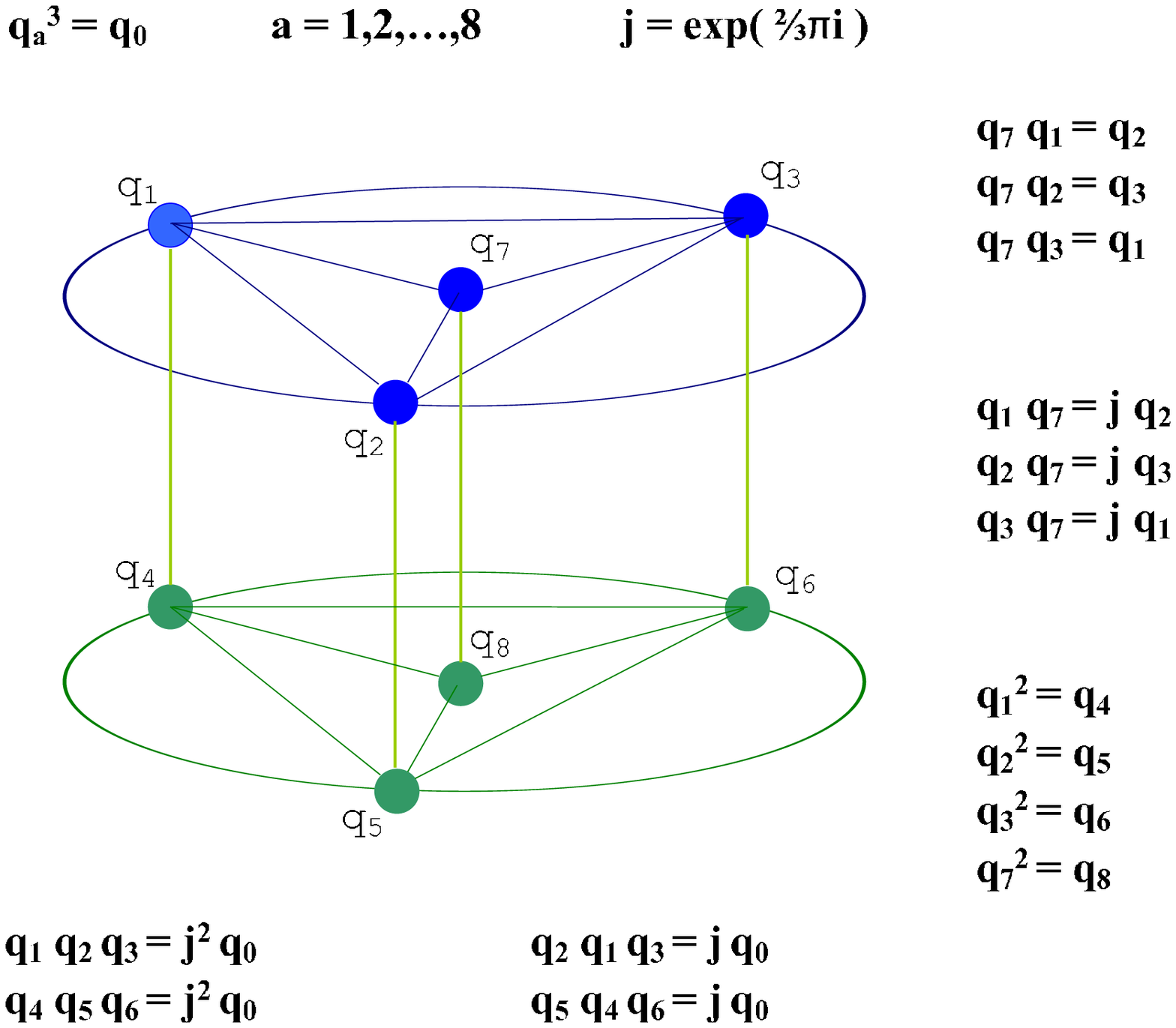}
        \label{fig:divalg3}
\end{figure}

The $q_k$ elements that satisfy the ternary algebra are:

\begin{eqnarray}
\{A,B,C\}_{S_3}=ABC+BCA+CAB-BAC-ACB-CBA.
\end{eqnarray}
Here $j=\exp(2{\bf i} \pi/3)$ and  
$S_3$ is the permutation group of three elements.

We can consider the $3\times 3$  matrix realization of $q-$-algebra:

\begin{eqnarray}
&&q_1=
\left (
\begin{array}{ccc}
_{0} &1&_{0} \\
_{0} &_{0} &1\\
1&_{0} &_{0} \\
\end{array}
\right),
\,
q_2= 
\left (
\begin{array}{ccc}
_{0}&1&_{0}  \\
_{0}&_{0} &j  \\
j^2&_{0} &_{0} \\
\end{array}
\right)
\,q_3=
\left (
\begin{array}{ccc}
_{0} &1&_{0}     \\
_{0} &_{0} &j^2  \\
j&_{0} &_{0}     \\
\end{array}
\right)
\nonumber\\
&&q_4=
\left (
\begin{array}{ccc}
_{0} &_{0} &1    \\
1&_{0} &_{0}   \\
_{0} &1&_{0}     \\
\end{array}
\right)
\,q_5=
\left (
\begin{array}{ccc}
_{0} &_{0} &j    \\
1&_{0} &_{0}   \\
_{0} &j^2&_{0}     \\
\end{array}
\right)
\,q_6=
\left (
\begin{array}{ccc}
_{0} &_{0} &j^2    \\
1&_{0} &_{0}   \\
_{0} &j&_{0}     \\
\end{array}
\right)
\nonumber\\
&&q_7=
\left (
\begin{array}{ccc}
1&_{0} &_{0}     \\
_{0} &j&_{0}   \\
_{0} &_{0} &j^2    \\
\end{array}
\right)
\,q_8=
\left (
\begin{array}{ccc}
1&_{0}   &_{0}     \\
_{0} &j^2&_{0}   \\
_{0} &_{0}   &j    \\
\end{array}
\right)
\,q_0=
\left (
\begin{array}{ccc}
1&_{0} &_{0}     \\
_{0} &1&_{0}   \\
_{0} &_{0} &1    \\
\end{array}
\right)
\nonumber\\
\end{eqnarray}
which satisfy the ternary algebra:

\begin{equation}
\{q_k,q_l,q_m\}_{S_3}=f_{klm}^n q_n.
\end{equation}
We can check that each triple commutator $\{q_k,q_l,q_m\}$
is defined by triple numbers, $\{klm\}$, with $k,l,m=0,1,2,...,8$,  
that it gives just one matrix $q_n$ with the corresponding coefficient
$f_{klm}^n$ given in Table \ref{COM}.

\begin{table}
\centering
\caption{ \it  The ternary commutation relations}
\label{COM}
\vspace{.05in}
\begin{tabular}{||c|c|c|||c|c|c|||c|c|c||}
\hline
$   N     $&$  \{klm\} \rightarrow \{n\}  $&$  f_{klm}^n  
$&$ N     $&$  \{klm\} \rightarrow  \{n\}  $&$  f_{klm}^n 
$&$ N     $&$  \{klm\} \rightarrow \{n\}  $&$  f_{klm}^n 
$\\ \hline\hline 
$   1     $&$  \{123\} \rightarrow \{0\}  $&$  3(j^2-j)       
$&$ 2     $&$  \{124\} \rightarrow \{2\}  $&$  j(1-j)     
$&$ 3     $&$  \{126\} \rightarrow \{3\}  $&$  2(j^2-j)  
$\\ \hline      
$   4     $&$  \{125\} \rightarrow \{1\}  $&$  j(1-j)       
$&$ 5     $&$  \{127\} \rightarrow \{5\}  $&$  2(1-j)     
$&$ 6     $&$  \{128\} \rightarrow \{4\}  $&$  2(j^2-1)  
$\\ \hline      
$   7     $&$  \{120\} \rightarrow \{6\}  $&$  (j^2-j)       
$&$ 8     $&$  \{134\} \rightarrow \{3\}  $&$  (j^2-j)     
$&$ 9     $&$  \{136\} \rightarrow \{1\}  $&$  (j^2-j)  
$\\ \hline      
$   10     $&$  \{135\} \rightarrow \{2\}  $&$  2(j-j^2)       
$&$ 11     $&$  \{137\} \rightarrow \{4\}  $&$  2(j-1)     
$&$ 12     $&$  \{138\} \rightarrow \{6\}  $&$  2(1-j^2)  
$\\ \hline      
$   13     $&$  \{130\} \rightarrow \{5\}  $&$  j(1-j)       
$&$ 14     $&$  \{146\} \rightarrow \{6\}  $&$  (j^2-j)     
$&$ 15     $&$  \{145\} \rightarrow \{5\}  $&$  (j-j^2)  
$\\ \hline      
$   16     $&$  \{147\} \rightarrow \{7\}  $&$  (j^2-j)       
$&$ 17     $&$  \{148\} \rightarrow \{8\}  $&$  (j-j^2)     
$&$ 18     $&$  \{140\} \rightarrow \tilde O  $&$    0  
$\\ \hline      
$   19     $&$  \{165\} \rightarrow \{4\}  $&$  2j(1-j)       
$&$ 20     $&$  \{167\} \rightarrow \{8\}  $&$  2(1-j^2)     
$&$ 21     $&$  \{168\} \rightarrow \{0\}  $&$  3(1-j^2)  
$\\ \hline      
$   22     $&$  \{160\} \rightarrow \{7\}  $&$  (1-j^2)       
$&$ 23     $&$  \{157\} \rightarrow \{0\}  $&$  3(1-j)     
$&$ 24     $&$  \{158\} \rightarrow \{7\}  $&$  2(1-j)  
$\\ \hline      
$   25     $&$  \{150\} \rightarrow \{8\}  $&$  (1-j)       
$&$ 26     $&$  \{178\} \rightarrow \{1\}  $&$  (j-j^2)     
$&$ 27     $&$  \{170\} \rightarrow \{2\}  $&$  (j-1)  
$\\ \hline      
$   28     $&$  \{180\} \rightarrow \{3\}  $&$  (j^2-1)       
$&$ 29     $&$  \{234\} \rightarrow \{1\}  $&$  2(j^2-j)     
$&$ 30     $&$  \{236\} \rightarrow \{2\}  $&$  (j-j^2)  
$\\ \hline      
$   31     $&$  \{235\} \rightarrow \{3\}  $&$  (j-j^2)       
$&$ 32     $&$  \{237\} \rightarrow \{6\}  $&$  2(1-j)     
$&$ 33     $&$  \{238\} \rightarrow \{5\}  $&$  2(j^2-1)  
$\\ \hline      
$   34     $&$  \{230\} \rightarrow \{4\}  $&$  (j^2-j)       
$&$ 35     $&$  \{246\} \rightarrow \{5\}  $&$  2(j-j^2)     
$&$ 36     $&$  \{245\} \rightarrow \{4\}  $&$  (j-j^2)  
$\\ \hline      
$   37     $&$  \{247\} \rightarrow \{8\}  $&$  2(1-j^2)       
$&$ 38     $&$  \{248\} \rightarrow \{0\}  $&$  3(1-j^2)     
$&$ 39     $&$  \{240\} \rightarrow \{7\}  $&$  (1-j^2)  
$\\ \hline      
$   40     $&$  \{265\} \rightarrow \{6\}  $&$  (j^2-j)       
$&$ 41     $&$  \{267\} \rightarrow \{0\}  $&$  3(1-j)     
$&$ 42     $&$  \{268\} \rightarrow \{7\}  $&$  2(1-j)  
$\\ \hline      
$   43     $&$  \{260\} \rightarrow \{8\}  $&$  (1-j)       
$&$ 44     $&$  \{257\} \rightarrow \{7\}  $&$  (j^2-j)     
$&$ 45     $&$  \{258\} \rightarrow \{8\}  $&$  (j-j^2)  
$\\ \hline           
$   46     $&$  \{250\} \rightarrow \tilde O  $&$ 0       
$&$ 47     $&$  \{278\} \rightarrow \{2\}  $&$  (j-j^2)     
$&$ 48     $&$  \{270\} \rightarrow \{3\}  $&$  (j-1)  
$\\ \hline      
$   49     $&$  \{280\} \rightarrow \{1\}  $&$  (j^2-1)       
$&$ 50     $&$  \{346\} \rightarrow \{4\}  $&$  (j^2-j)     
$&$ 51     $&$  \{345\} \rightarrow \{6\}  $&$  2(j^2-j)  
$\\ \hline      
$   52     $&$  \{347\} \rightarrow \{0\}  $&$  3(1-j)      
$&$ 53     $&$  \{348\} \rightarrow \{7\}  $&$  2(1-j)     
$&$ 54     $&$  \{340\} \rightarrow \{8\}  $&$  (1-j)  
$\\ \hline      
$   55     $&$  \{365\} \rightarrow \{5\}  $&$  j^2-j       
$&$ 56     $&$  \{367\} \rightarrow \{7\}  $&$  (j^2-j)     
$&$ 57     $&$  \{368\} \rightarrow \{8\}  $&$  (j^2-j)  
$\\ \hline      
$   58     $&$  \{360\} \rightarrow \tilde O $&$  0       
$&$ 59     $&$  \{357\} \rightarrow \{8\}  $&$  2(1-j^2)     
$&$ 60     $&$  \{358\} \rightarrow \{0\}  $&$  3(1-j^2)  
$\\ \hline      
$   61     $&$  \{350\} \rightarrow \{7\} $&$   (1-j^2)       
$&$ 62     $&$  \{378\} \rightarrow \{3\}  $&$  (j-j^2)     
$&$ 63     $&$  \{370\} \rightarrow \{1\}  $&$  (j-1)  
$\\ \hline      
$   64     $&$  \{380\} \rightarrow \{2\} $&$   (j^2-1)       
$&$ 65     $&$  \{465\} \rightarrow \{0\}  $&$  3(j-j^2)     
$&$ 66     $&$  \{467\} \rightarrow \{3\}  $&$  2(j-1)  
$\\ \hline      
$   67     $&$  \{468\} \rightarrow \{1\}  $&$   2(1-j^2)       
$&$ 68     $&$  \{460\} \rightarrow \{2\}  $&$  (j-j^2)     
$&$ 69     $&$  \{457\} \rightarrow \{1\}  $&$  2(1-j)  
$\\ \hline      
$   70     $&$  \{458\} \rightarrow \{2\} $&$   2(j^2-1)       
$&$ 71     $&$  \{450\} \rightarrow \{3\}  $&$  (j^2-j)     
$&$ 72     $&$  \{478\} \rightarrow \{4\}  $&$  (j^2-j)  
$\\ \hline      
$   73     $&$  \{470\} \rightarrow \{6\} $&$   (1-j)       
$&$ 74     $&$  \{480\} \rightarrow \{5\}  $&$  (1-j^2)     
$&$ 75     $&$  \{657\} \rightarrow \{2\}  $&$  2(j-1)  
$\\ \hline      
$   76     $&$  \{658\} \rightarrow \{3\} $&$   2(1-j^2)       
$&$ 77     $&$  \{650\} \rightarrow \{1\}  $&$  (j-j^2)     
$&$ 78     $&$  \{678\} \rightarrow \{6\}  $&$  (j^2-j)  
$\\ \hline      
$   79     $&$  \{670\} \rightarrow \{5\} $&$   (1-j)       
$&$ 80     $&$  \{680\} \rightarrow \{4\}  $&$  (1-j^2)     
$&$ 81     $&$  \{578\} \rightarrow \{5\}  $&$  (j^2-j)  
$\\ \hline      
$   82     $&$  \{570\} \rightarrow \{4\} $&$   (1-j)       
$&$ 83     $&$  \{580\} \rightarrow \{6\}  $&$  (1-j^2)     
$&$ 84     $&$  \{780\} \rightarrow \tilde O  $&$  0  
$\\ \hline      
\end{tabular}
\end{table}

There are $C_9^2=84$ ternary commutation relations, but there are 
also $C_8^2=28$ commutation relations which
correspond to the $su(3)$ algebra! Therefore, it is natural 
to represent the $q$-numbers as a ternary generalization of quaternions. 
The $S_3$ commutation relations  naturally go to the binary, $S_2$, Lie commutation
relations:

\begin{equation}
\{q_a,q_b, q_0\}_{S_3}=q_aq_bq_0+q_bq_0q_a+q_0q_aq_b-q_bq_aq_0
-q_aq_0q_b-q_0q_bq_a=q_aq_b - q_bq_a,
\end{equation}
where $a \neq b \neq 0$. 
On table for those 28 cases, one can see that we have $\{kl0\}$.

Let's review ternary-quaternions in matrix form:
\begin{eqnarray}
Q=q_ax_a=
\left (
\begin{array}{ccc}
z_0                 &   z_1                &   \tilde{z}_2 \\
z_2                 & \tilde{z}_0          &   \tilde{z}_1 \\
\tilde{\tilde{z}}_1 & \tilde{\tilde{z}}_2  &   \tilde{\tilde{z}}_0 \\
\end{array}
\right),
\end{eqnarray}
where 
\begin{eqnarray}
\begin{array}{ccc}
z_0=x_0+x_7+x_8,                         & 
\tilde {z}_0=x_0+j x_7+j^2 x_8,           &
\tilde {\tilde {z}}_0=x_0+j^2 x_7+j x_8, \\
z_1=x_1+x_2+x_3,& 
\tilde {z}_1=x_1+j x_2+j^2 x_3, &
\tilde {\tilde {z}}_1=x_1+j^2 x_2+j x_3, \\
z_2=x_4+x_5+x_6,&\tilde {z}_2=x_4+j^2 x_4+j x_5, &
\tilde {\tilde {z}}_2=x_4+j  x_5+j^2 x_6.  \\
\end{array}
\end{eqnarray}

According to the permutation groups $Z_2$ and  $Z_3$,
 one can consider the $Z_2$ and the $Z_3$ transposition operations:
\begin{eqnarray}
\begin{array}{cc}
t(Z_2):           &  \qquad 1 \rightarrow 2;  \,2 \rightarrow 1 \\
T_+(Z_3):  &  \qquad 1 \rightarrow 2 ; \,2\rightarrow 3;\, 3 \rightarrow 1\\
T_-(Z_3):   &  \qquad 3 \rightarrow 2 ; \,2\rightarrow 1;\, 1 \rightarrow 3. \\
\end{array}
\end{eqnarray}

For example, in the  $Z_3$ transposition, we can obtain
the following transformation (conjugations) 
for the $q_i$ ($i=1,2,3$) elements:
\begin{eqnarray}
&&q_1^{T_+}=q_1,\qquad q_2^{T_+}=j^2q_2,\qquad q_3^{T_+}=jq_3,\nonumber\\ 
&&q_4^{T_+}=q_4,\qquad q_5^{T_+}=j^2q_5,\qquad q_6^{T_+}=jq_6,\nonumber\\ 
&&q_0^{T_+}=q_0,\qquad q_7^{T_+}=j^2q_7,\qquad q_8^{T_+}=jq_8.\nonumber\\ 
\end{eqnarray}
and  
\begin{eqnarray}
&&q_1^{T_-}=q_1,\qquad q_2^{T_-}=jq_2,\qquad q_3^{T_-}=j^2q_3\nonumber\\ 
&&q_4^{T_-}=q_4,\qquad q_5^{T_-}=jq_5,\qquad q_6^{T_-}=j^2q_6\nonumber\\ 
&&q_0^{T_-}=q_0,\qquad q_7^{T_-}=jq_7,\qquad q_8^{T_-}=j^2q_8, \nonumber\\
\end{eqnarray}
respectively.

In the two subsequent ${Z_3}^+$ transpositions,
the $Q$-matrices are transformed in the following way:

\begin{eqnarray}
Q^{T_+}=
\left(
\begin{array}{ccc}
x_0 +j^2x_7+j  x_8  & x_1+j^2x_2 +j   x_3  & x_4 +j   x_5 +j^2 x_6 \\
x_4 +j^2x_5+j  x_6  & x_0+   x_7 +    x_8  & x_1 +    x_2 +    x_3 \\
x_1 +j  x_2+j^2x_3  & x_4+   x_5 +    x_6  & x_0 +j   x_7 +j^2 x_8 \\
\end{array}
\right),
\end{eqnarray}
and
\begin{eqnarray}
Q^{TT_+}=
\left (
\begin{array}{ccc}
x_0 +j  x_7+j^2 x_8 & x_1+j  x_2 +j^2 x_3  & x_4 +    x_5 +    x_6 \\
x_4 +j^2x_5+j   x_6 & x_0+j^2x_7 +j   x_8  & x_1 +j^2 x_2 +j   x_3 \\
x_1 +   x_2+    x_3 & x_4+j^2x_5 +j   x_6  & x_0 +    x_7 +    x_8 \\
\end{array}
\right).
\end{eqnarray}

In this  ternary conjugation, the cubic form $QQ^TQ^{TT}$,

\begin{eqnarray}
&&      (x_0^3+x_7^3+x_8^3-3x_0x_7x_8)
+       (x_1^3+x_2^3+x_3^3-3x_1x_2x_3)
+       (x_4^3+x_5^3+x_6^3-3x_4x_5x_6)        \nonumber\\
&&-3 x_0(x1x_4+x_2x_5+x_3x_6)
-3j^2x_7(x_1x_5+x_2x_6+x_3x_4)
-3j  x_8(x_1x_6+x_2x_4+x_3x_5),                \nonumber\\
\end{eqnarray} 
is "binry" complex. 
This form is also symmetric under the complex 
conjugation and the following transformations:
$x_1 \leftrightarrow x_4,  x_2 \leftrightarrow x_5,
x_3 \leftrightarrow x_6, x_7 \leftrightarrow x_8$.
One can take in account the binary complex conjugation to define the following norm:
\begin{equation}
<Q>=(QQ^TQ^{TT})(QQ^TQ^{TT})^t
\end{equation}, where the $T$ is $T_+$ or $T_-$.  
This norm can result in an appearance of group properties of ternary quaternions $<Q>=1$.

If the binary alternative division algebras (real numbers, complex numbers,
quaternions, octonions) over the real numbers
have the dimensions           $2^n$, $n=0,1,2,3,4,...$, the ternary  algebras 
have the following dimensions $3^n$, $n=0,1,2,3,4,...$, respectively:

\begin{eqnarray}
\begin{array}{cc|cc}
 {\mathbb R}    :& \, 2^0  \,=\,1       
&{\mathbb R}    :& \, 3^0  \,=\,1                       \\
 {\mathbb C}    :& \, 2^1  \,=\,1+1     
&{\mathbb TC}   :& \, 3^1  \,=\,1+1+1                   \\     
 {\mathbb H}    :& \, 2^2  \,=\,1+2+1   
&{\mathbb TH}   :& \, 3^2  \,=\,1+2+3+2+1               \\      
 {\mathbb O}    :& \, 2^3  \,=\,1+3+3+1 
&{\mathbb TO}   :& \, 3^3  \,=\,1+3+6+7+6+3+1           \\
 {\mathbb S}    :& \, 2^4  \,=\,1+4+6+4+1 
&{\mathbb TS}   :& \, 3^4  \,=\,1+4+10+16+19+16+10+4+1  \\
\end{array}
\end{eqnarray}  

In the last line the sedenions do not produce the division algebra.
For both cases we have the unit elements $e_0$ and $q_0$, and 
the $(n-1)$ basis elements

\begin{eqnarray} 
\begin{array}{cc}
\{e_a: & e_a^2=-e_0 \}, \\
\{q_a: & q_a^3= q_0 \}, \\
\end{array}
\end{eqnarray}
respectively.  

\begin{eqnarray}
{\mathbb R}  \rightarrow {\mathbb TC}  \rightarrow {\mathbb TH} \rightarrow 
{\mathbb TO} \rightarrow {\mathbb TS}  \rightarrow   \ldots
\end{eqnarray}

To build the ternary "octonions" from  ternary "quaternion"' 
one needs the additional basis element. For illustration we take the following 
three basis elements: the previous two, $q_1,q_7$ and new third element $q_{21}$. Then, 
applying the generalized Cayley--Dickson method, we can get the $27$-dimensional
algebra with $q_a^3=q_0$, $a=1,2,...,26$ (see figure \ref{fig:octonion27}) :
\begin{eqnarray}
TO=
&=&(x_0q_0+x_1q_1+x_2q_2+x_3q_3+x_4q_4+x_5q_5+x_6q_6 +x_7q_7+x_8q_8)            \nonumber\\
&+&(y_0q_0+y_1q_1+y_2q_2+y_3q_3+y_4q_4+y_5q_5+y_6q_6 +y_7q_7+y_8q_8) q_{21}     \nonumber\\
&+&(z_0q_0+z_1q_1+z_2q_2+z_3q_3+z_4q_4+z_5q_5+z_6q_6 +z_7q_7+z_8q_8) q_{21}^2,  \nonumber\\
\end{eqnarray}

\section{Appendix: $x^3+y^3+u^3-3xyu=1$.}

\begin{eqnarray}
\rho &=&x+y+u   \nonumber\\
z&=&x+jy+j^2u   \nonumber\\
\bar z&=&x+j^2y+ju \nonumber\\
\end{eqnarray}
(We use the sign $\bar z$ to define binary  conjugation of
ordinary complex numbers, and sign $\tilde z$ ternary complex
conjugation.)   Using these new variables one can rewrite the
cubic equation
\begin{eqnarray}
x^3+y^3+u^3-3xyu=1
\end{eqnarray}
in the following form
\begin{eqnarray}
\rho |z|^2=1.
\end{eqnarray}
Let  substitute the $\rho $ in this equation:

\begin{eqnarray}
z \bar z &=& (x+jy+j^2u)(x^2+j^2y+j u)=(x^2+y^2+u^2)-(xy+xu+yu) \nonumber\\
&=&(x-\frac{y+u}{2})^2+\frac{3}{4}(y^2+u^2)-\frac{3}{2}yu=\nonumber\\
&=&(\rho-\frac{3}{2}(y+u))^2+\frac{3}{4}(y-u)^2
\nonumber\\
&=&\frac{1}{\rho}, \nonumber\\
\end{eqnarray}
or
\begin{eqnarray}
(\rho-\frac{3}{2}(y+u))^2+(\frac{\sqrt{3}}{2}(y-u))^2=(\frac{1}{\sqrt{\rho}})^2.
\end{eqnarray}

Take new variables
\begin{eqnarray}
\frac{3}{2}(y+u)-\rho&=&\frac{1}{\sqrt{\rho}} \cos \phi  \nonumber\\
\frac{\sqrt{3}}{2}(y-u)&=&\frac{1}{\sqrt{\rho}} \sin \phi.
\nonumber\\
\end{eqnarray}

In result we can get the next parametrization

\begin{eqnarray}
x&=&\frac{\rho}{3}-\frac{2}{3\sqrt{\rho}}\cos \phi \nonumber\\
y&=&\frac{\rho}{3}+\frac{1}{3\sqrt{\rho}}(\cos \phi +\sqrt{3}\sin \phi) \nonumber\\
u&=&\frac{\rho}{3}+\frac{1}{3\sqrt{\rho}}(\cos \phi -\sqrt{3}\sin \phi) \nonumber\\
\end{eqnarray}
or
\begin{eqnarray}
x&=&\frac{\rho }{3}-\frac{2}{3\sqrt{\rho}}\cos \phi \nonumber\\
y&=&\frac{\rho }{3}+\frac{2}{3\sqrt{\rho}}\sin (\phi + \pi/6) \nonumber\\
u&=&\frac{\rho }{3}+\frac{2}{3\sqrt{\rho }}\sin (\phi - \pi/6) \nonumber\\
\end{eqnarray}

Note, that
\begin{eqnarray}
\rho >0, \qquad   0 \le  \phi \le 2\pi.
\end{eqnarray}

\begin{figure}
        \centering
                \includegraphics[type=eps,ext=.eps,read=.eps]{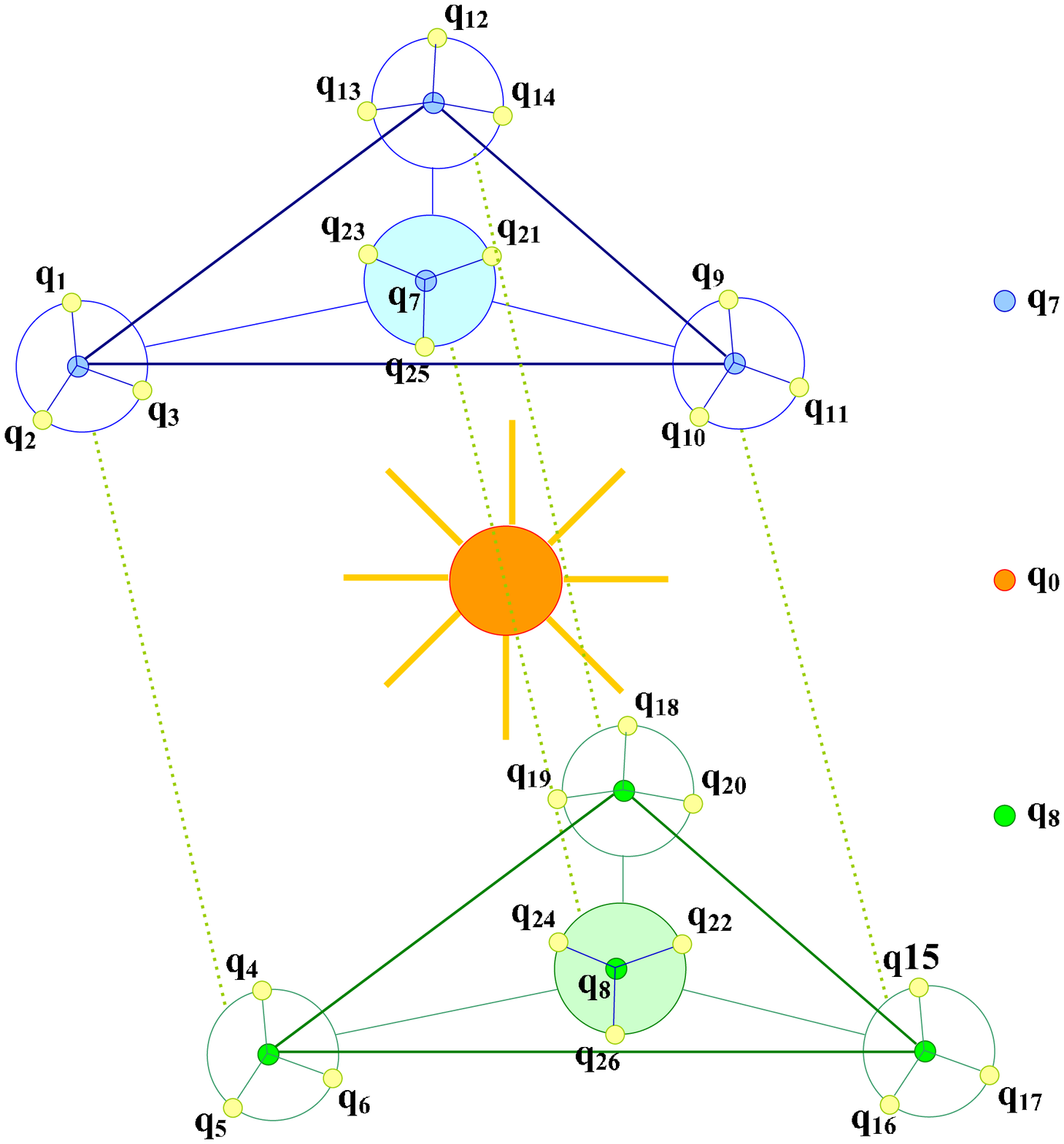}
        \label{fig:octonion27}
\end{figure}

\section{Ternary algebras of higher rank}

Now we would like to create the ternary algebra of rank 3 from
the simple ternary algebra, which can be described by a tetrahedron
Berger graph \cite{V}.  This is similar to the way of creating the
$su(3)$ ($su(r+1)$) Lie algebra from $su(2)$-algebras.  
The $su(r+1)$ algebra is
the algebra
of rank r having   $r$-simple roots $\alpha_a$ ($a=1,2,...,r$),
each of them  corresponding to the $su(2)/u(1)$ algebra.
This is a way to construct the $su(r+1)$ algebra from r-$su(2)$
algebras!
We can try to do same to construct the ${\rm Berg}_r$ algebras from
the minimal simple ${\rm Berg}$-algebra.

\begin{figure}[htpb]
\begin{center}
\mbox{\epsfig{file=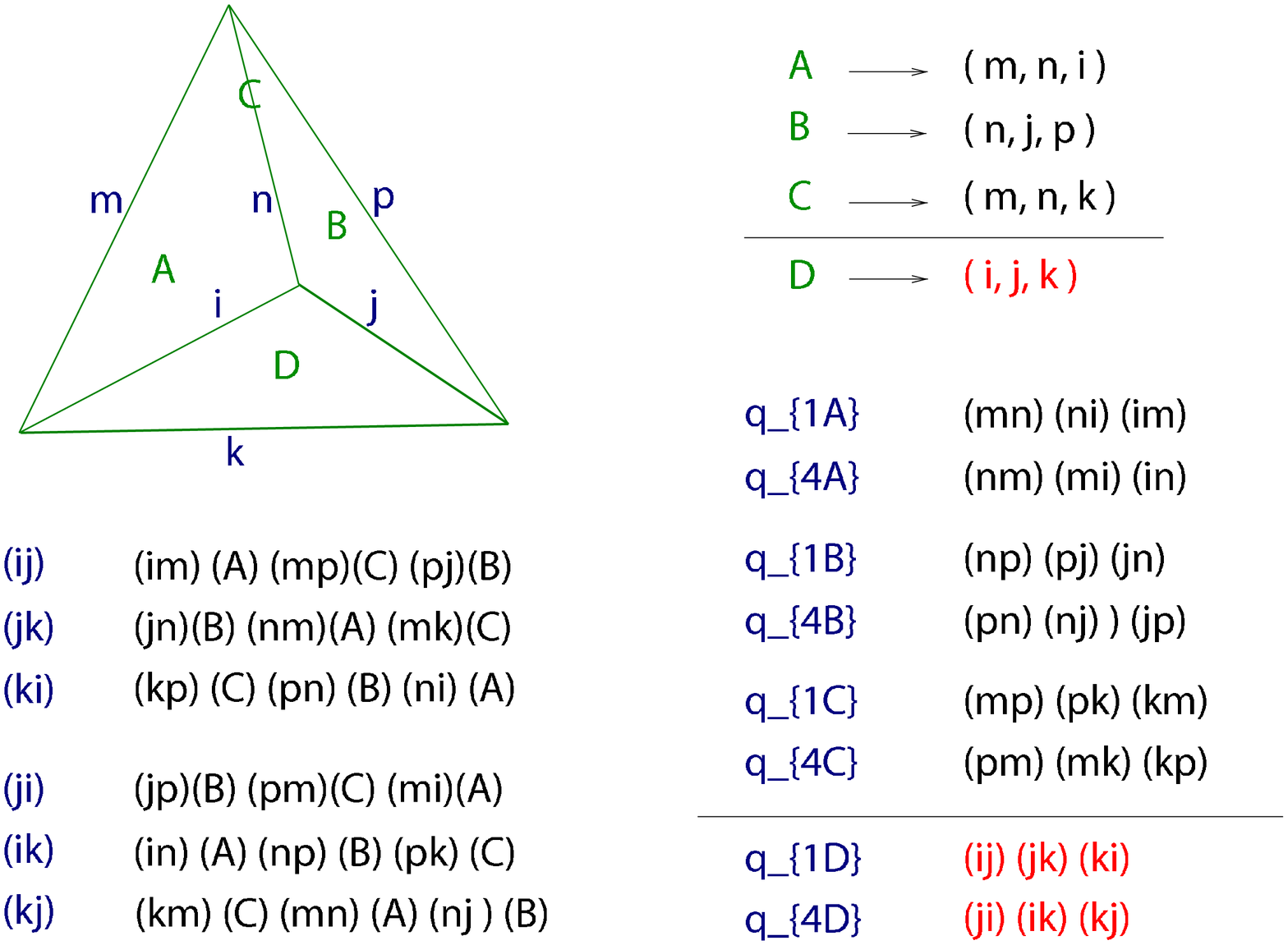,height=10cm,width=14cm}}
\end{center}
\caption{The rank-3 Berger graph}
\label{fig:Brank3}
\end{figure}

Let us take  three ternary algebras based on the 
 $q_A(m,n,i)$,$q_B(n,p,j)$, $q_C(m,p,k)$ matrices, where
$A(m,n,p),B(n,p,j), C(m,p,k)=0,1,2,...,8$.
According to our conception (Figure \ref{fig:Brank3}) we would like to obtain  the ternary 
algebra of higher rank defined by generators $q_D(i,j,k)$. For this we should 
consider all  ternary commutation relations: 
$[Q_a, Q_a', Q_a'']_{S_3}$, where $Q_a$ is the complete set of
all considered elements $Q_a=\{q_A,q_B,q_C\}$.
For example, we can consider the following:

\small 
\begin{eqnarray}
q_{A}=
\left (
\begin{array}{cccccc}
_{0} &mn& _{0} & _{0}& _{0}& _{0} \\
_{0} & _{0} &ni& _{0}& _{0}& _{0} \\
im& _{0} & _{0} & _{0}& _{0}& _{0} \\
_{0} & _{0} & _{0} & _{0}& _{0}& _{0} \\
_{0} & _{0} & _{0} & _{0}& _{0}& _{0} \\
_{0} & _{0} & _{0} & _{0}& _{0}& _{0} \\
\end{array}
\right)\,
q_{TA}=
\left (
\begin{array}{cccccc}
_{0}&_{0}&_{0}&mi&_{0}&_{0} \\
nm&_{0}&_{0}&_{0}&_{0}&_{0} \\
_{0}&in&_{0}&_{0}&_{0}&_{0} \\
_{0}&_{0}&_{0}&_{0}&_{0}&_{0} \\
_{0}&_{0}&_{0}&_{0}&_{0}&_{0} \\
_{0}&_{0}&_{0}&_{0}&_{0}&_{0} \\
\end{array}
\right)\,
q_{A}^0=
\left (
\begin{array}{cccccc}
mm&_{0}&_{0}&_{0}&_{0}&_{0} \\
_{0}&nn&_{0}&_{0}&_{0}&_{0} \\
_{0}&_{0}&ii&_{0}&_{0}&_{0} \\
_{0}&_{0}&_{0}&_{0}&_{0}&_{0} \\
_{0}&_{0}&_{0}&_{0}&_{0}&_{0} \\
_{0}&_{0}&_{0}&_{0}&_{0}&_{0} \\
\end{array}
\right)
\nonumber\\
\end{eqnarray}

\begin{eqnarray}
q_{B}=
\left (
\begin{array}{cccccc}
_{0} &_{0} &_{0} &_{0}&_{0}&_{0} \\
_{0} &_{0} &_{0} &np& ._{0}&_{0} \\
_{0} &_{0} &_{0} &_{0}&_{0}&_{0} \\
_{0} &_{0} &_{0} &_{0}&_{0}&pj\\
_{0} &_{0} &_{0} &_{0}&_{0}&_{0} \\
0  &jn &_{0} &_{0}&_{0}&_{0} \\
\end{array}
\right)\,
q_{B}^T=
\left (
\begin{array}{cccccc}
_{0} &_{0} &_{0}&_{0} &_{0}&_{0}  \\
_{0} &_{0} &_{0}&_{0} &_{0}&nj \\
_{0} &_{0} &_{0}&_{0} &_{0}&_{0}  \\
_{0} &pn&_{0}&_{0} &_{0}&_{0}  \\
_{0} &_{0} &_{0}&_{0} &_{0}&_{0}  \\
_{0} &_{0} &_{0}&jp&_{0}&_{0}  \\
\end{array}
\right)\,
q_{B}^0=
\left (
\begin{array}{cccccc}
_{0} &_{0} &_{0} &_{0}&_{0}&_{0} \\
_{0} &nn&_{0} &_{0}&_{0}&_{0} \\
_{0} &_{0} &_{0} &_{0}&_{0}&_{0} \\
_{0} &_{0} &_{0} &pp& ._{0}&_{0} \\
_{0} &_{0} &_{0} &_{0}&_{0}&_{0} \\
_{0} &_{0} &_{0} &_{0}&_{0}&jj \\
\end{array}
\right)
\nonumber\\
\end{eqnarray}

\begin{eqnarray}
q_{C}=
\left (
\begin{array}{cccccc}
_{0} &_{0} &_{0} &mp&_{0} &_{0} \\
_{0} &_{0}&_{0} &_{0} &_{0} &_{0} \\
_{0} &_{0}&_{0} &_{0} &_{0} &_{0} \\
_{0} &_{0}&_{0} &_{0} &pk&_{0} \\
km&_{0}&_{0} &_{0} &_{0} &_{0} \\
_{0} &_{0}&_{0} &_{0} &_{0} &_{0} \\
\end{array}
\right)\,
q_{C}^T=
\left (
\begin{array}{cccccc}
_{0} &_{0} &_{0}&_{0} &mk&_{0}  \\
_{0} &_{0} &_{0}&_{0} &_{0} &_{0}  \\
_{0} &_{0} &_{0}&_{0} &_{0} &_{0}  \\
pm&_{0} &_{0}&_{0} &_{0} &_{0}  \\
_{0} &_{0} &_{0}&kp&_{0} &_{0}  \\
_{0} &_{0} &_{0}&_{0} &_{0} &_{0}  \\
\end{array}
\right)\,
q_{C}^0=
\left (
\begin{array}{cccccc}
mm&_{0} &_{0} &_{0} &_{0} &_{0} \\
_{0} &_{0} &_{0} &_{0} &_{0} &_{0} \\
_{0} &_{0} &_{0} &_{0} &_{0} &_{0} \\
_{0} &_{0} &_{0} &pp&_{0} &_{0} \\
_{0} &_{0} &_{0} &_{0} &kk&_{0} \\
_{0} &_{0} &_{0} &_{0} &_{0} &_{0} \\
\end{array}
\right)
\nonumber\\
\end{eqnarray}

\begin{eqnarray}
q_{D}=
\left (
\begin{array}{cccccc}
_{0} &_{0} &_{0} &_{0}&_{0} &_{0}  \\
_{0} &_{0} &_{0} &_{0}&_{0} &_{0}  \\
_{0} &_{0} &_{0} &_{0}&ik&_{0}  \\
_{0} &_{0} &_{0} &_{0}&_{0} &_{0}  \\
_{0} &_{0} &_{0} &_{0}&_{0} &kj \\
_{0} &_{0} &ji&_{0}&_{0} &_{0}  \\
\end{array}
\right)\,
q_{D}^T=
\left (
\begin{array}{cccccc}
_{0} &_{0} &_{0} &_{0} &_{0} &_{0}  \\
_{0} &_{0} &_{0} &_{0} &_{0} &_{0} \\
_{0} &_{0} &_{0} &_{0} &_{0} &ij  \\
_{0} &_{0} &ki&_{0} &_{0} &_{0}  \\
_{0} &_{0} &_{0} &_{0} &_{0} &_{0}  \\
_{0} &_{0} &_{0} &_{0} &jk&_{0}  \\
\end{array}
\right)\,
q_{D}^0=
\left (
\begin{array}{cccccc}
_{0} &_{0} &_{0} &_{0}&_{0} &_{0} \\
_{0} &_{0} &_{0} &_{0}&_{0} &_{0} \\
_{0} &_{0} &ii&_{0}&_{0} &_{0} \\
_{0} &_{0} &_{0} &_{0}&_{0} &_{0} \\
_{0} &_{0} &_{0} &_{0}&kk&_{0} \\
_{0} &_{0} &_{0} &_{0}&_{0} &jj \\
\end{array}
\right)
\nonumber\\
\end{eqnarray}

\begin{eqnarray}
t_{su(2)}=
\left (
\begin{array}{c|c|cc|c|c}
mm &_{0}  &_{0}  &_{0} &_{0}   &mj  \\\hline
_{0}&nn &_{0}  &_{0} &nk  &_{0}   \\ \hline
_{0}&_{0}  &ii &pi&_{0}   &_{0}   \\
_{0}&_{0}  &ip &pp&_{0}   &_{0}   \\ \hline
_{0}&kn &_{0}  &_{0} &kk  &kj  \\ \hline
jm &_{0}  &_{0}  &_{0} &_{0}   &jj  \\
\end{array}
\right)
\nonumber\\
\end{eqnarray}

\normalsize

\section{Appendix: Quaternary algebra}
Now we would like to  illustrate the quaternary algebra.
\begin{eqnarray}
q_m^4=1, \,\,\,m=0,1,2,3,...,14,15.\nonumber\\
\end{eqnarray}

\begin{eqnarray}
&& q_{m+1}=q_{13}^{m} q_1, \,\,q_{4-m}=q_{15}^mq_4,\,\, 
m=0,1,2,3\nonumber\\
&& q_{5+m}=q_{13}^{m} q_5, \,\,q_{8-m}=q_{15}^mq_8,\,\, 
m=0,1,2,3,\nonumber\\
&& q_{9+m}=q_{13}^{m} q_9, \,\,q_{12-m}=q_{15}^mq_{12}.\,\, 
m=0,1,2,3,\nonumber\\
\end{eqnarray}

\begin{eqnarray}
&&q_{13}^2=q_{14},\,\,\, q_{13}^3=q_{15}, \nonumber\\
&&q_{14}^2=q_0,\nonumber\\
&&q_{15}^2=q_{14},\,\,\, q_{15}^3=q_{13}\nonumber\\
\end{eqnarray}

\begin{eqnarray}
&& q_1^2=q_5, \,\,\,q_2^2=jq_7,\,\,\,q_3^2=j^2q_5,\,\,\,q_4^2=j^3q_7,
\nonumber\\
&& q_1^3=q_9,\,\,\, q_2^3=j^3q_{12},\,\,\,q_3^3=j^2q_{11},
\,\,\,q_4^3=jq_{10}\nonumber\\
\end{eqnarray}

\begin{eqnarray}
q_5^2=q_0,\,\,\,q_6^2=j^2q_{14}^2,\,\, q_{7}^2=q_0, \,\,q_8^2=j^2q_{14}.
\end{eqnarray}

\begin{eqnarray}
&& q_{9}^2=q_5, \,\,\,q_{10}^2=j^3q_7,\,\,\,
q_{11}^2=j^2q_5,\,\,\,q_{12}^2=jq_7,
\nonumber\\
&& q_{9}^3=q_1,\,\,\, q_{10}^3=jq_{4},\,\,\,
q_{11}^3=j^2q_{3},\,\,\,q_{12}^3=j^3q_{2}\nonumber\\
\end{eqnarray}

We can consider the $4\times 4$  matrix realization of $q$-algebra:
\begin{eqnarray}
&&q_1=
\left (
\begin{array}{cccc}
_{0}&1&_{0} &_{0} \\
_{0}&_{0} &1&_{0} \\
_{0}&_{0} &_{0} &1\\
1&_{0} &_{0} &_{0} \\
\end{array}
\right)
\,
q_2= 
\left (
\begin{array}{cccc}
_{0}&1&_{0} &_{0}   \\
_{0}&_{0} &j&_{0}   \\
_{0}&_{0} &_{0} &j^2 \\
j^3&_{0} &_{0} &_{0}   \\
\end{array}
\right)
\,q_3=
\left (
\begin{array}{cccc}
_{0}&1&_{0}  &_{0}   \\
_{0}&_{0} &j^2& 0  \\
_{0}&_{0} &_{0}  &1  \\
j^2& _{0}&_{0}  &_{0}   \\
\end{array}
\right)
q_4=
\left (
\begin{array}{cccc}
_{0}&1&_{0}   &_{0}    \\
_{0}&_{0} &j^3&_{0}    \\
_{0}&_{0} &_{0}   &j^2 \\
j&_{0} &_{0}   &_{0}     \\
\end{array}
\right)
\nonumber\\
&&q_5=
\left (
\begin{array}{cccc}
_{0}&_{0} &1&_{0} \\
_{0}&_{0} &_{0} &1\\
1&_{0} &_{0} &_{0} \\
_{0}&1&_{0} &_{0} \\
\end{array}
\right)
\,
q_6= 
\left (
\begin{array}{cccc}
_{0}&_{0}   &1&_{0}   \\
_{0}&_{0}   &_{0} &j  \\
j^2  &_{0}   &_{0} &_{0}   \\
_{0}&j^3&_{0} &_{0}   \\
\end{array}
\right)
\,
q_7=
\left (
\begin{array}{cccc}
_{0}&_{0} &1 &_{0}   \\
_{0}&_{0} &_{0}  &j^2\\
1 &_{0} &_{0}  &_{0}   \\
_{0}&j^2& _{0}&_{0}   \\
\end{array}
\right)
q_8=
\left (
\begin{array}{cccc}
_{0}&_{0} &1&_{0}     \\
_{0}&_{0} &_{0}  &j^3   \\
j^2&_{0} &_{0} &_{0}     \\
_{0}&j&_{0} &_{0}     \\
\end{array}
\right)
\nonumber\\
&&q_9=
\left (
\begin{array}{cccc}
_{0}&_{0} &_{0} &1\\
1&_{0} &_{0} &_{0} \\
_{0}&1&_{0} &_{0} \\
_{0}&_{0} &1&_{0} \\
\end{array}
\right)
\,
q_{10}= 
\left (
\begin{array}{cccc}
_{0}&_{0}     &_{0}   &1  \\
j    &_{0}     &_{0}   &_{0}   \\
_{0}&j^2  &_{0}   &_{0}   \\
_{0}&_{0}     &j^3&_{0}   \\
\end{array}
\right)
\,q_{11}=
\left (
\begin{array}{cccc}
0   &_{0} &_{0}    &1  \\
j^2 &_{0} &_{0}    &_{0}   \\
0   &1&_{0}    &_{0}   \\
0   &_{0} &j^2 &_{0}   \\
\end{array}
\right)
q_{12}=
\left (
\begin{array}{cccc}
_{0}&_{0}   &_{0} &1    \\
j^3  &_{0}   &_{0} &_{0}    \\
_{0}&j^2&_{0} &_{0}     \\
_{0}&_{0}   &j&_{0}     \\
\end{array}
\right)
\nonumber\\
&&q_{13}=
\left (
\begin{array}{cccc}
1&_{0} &_{0}   &_{0}   \\
_{0}&j&_{0}   &_{0}   \\
_{0}&_{0} &j^2&_{0}   \\
_{0}&_{0} &_{0}   &j^3\\
\end{array}
\right)
\,
q_{14}= 
\left (
\begin{array}{cccc}
1    &_{0}     &_{0}   &_{0}   \\
_{0}&j^2  &_{0}   &_{0}   \\
_{0}&_{0}     &1  &_{0}   \\
_{0}&_{0}     &_{0}   &j^2  \\
\end{array}
\right)
\,
q_{15}=
\left (
\begin{array}{cccc}
1&_{0}  &_{0}    &_{0}   \\
_{0}&j^3& 0   &_{0}   \\
_{0}&_{0}  &j^2 &_{0}   \\
_{0}&_{0}  &_{0}    &j  \\
\end{array}
\right)
q_{0}=
\left (
\begin{array}{cccc}
1&_{0} &_{0} &_{0}   \\
_{0}&1&_{0} &_{0}   \\
_{0}&_{0} &1&_{0}   \\
_{0}&_{0} &_{0} &1  \\
\end{array}
\right)
\nonumber\\
\end{eqnarray}
where $j=\exp{ 2 {\bf i} \pi/4}$.

\newpage

\section{Conclusions and Acknowledgements}
Our interest in the search for the $n$-ary algebras has a long history and 
started in 1998 when we have read  the pioneer
article of  P. Candelas and M. Font \cite{Font}.
 
We found the ''minimal'' non-Abelian ternary algebra which we intend 
using for solving  the  set of  Berger graphs. This could help us to 
find the ternary generalization of Cartan--Killing--Lie classification.
Also it is very important to 
understand the ternary "octonions'' and find a link between binary 
and ternary exceptional graphs and algebras.
Examples of the ternary algebras, which are presented here, could
help us to find some physical applications, for example, in solving of 
the tetrahedron Baxter equation. The other physical application
is related to searching for a ternary generalization of Lorentz group.
In the end we hope that the ternary symmetries could help us to make 
the quantium theory of membranes.

We would like to express our thanks for many discussions, advices 
and support to
U. Aglietti, E. Alvarez, L. Alvarez-Gaume, I. Antoniadis,
P. Auranche,  I.Bakas, G. Belanger,
 P. Chankowski, G. Costa, J.Ellis,  L. Fellin, C. Gomez,   L. Lipatov, 
V. Maroussov, P. Sorba, A. Sabio-Vera, J-B. Zuber.
 
We thank Prof. G. Girardi and Prof. R. Kerner  very much for reading this article and 
making useful comments.

\end{document}